\documentclass[12pt]{article}
\begin{document}

\begin{center}
{\large \bf Some thoughts on an information-theoretic motivated nonlinear Schrodinger equation}
\end{center}
\vspace{0.1in}

\begin{center}

{Rajesh R. Parwani\footnote{Email: parwani@nus.edu.sg} }

\vspace{0.3in} 

{Department of Physics and University Scholars Programme,\\}
{National University of Singapore,\\}
{Kent Ridge,\\}
{ Singapore.}

\vspace{0.3in}
5 October 2006
\end{center}
\vspace{0.1in}
\begin{abstract}

I begin by reviewing the arguments leading to a nonlinear generalisation of Schrodinger's equation within the context of the maximum uncertainty principle. Some exact and perturbative properties of that equation are then summarised: those results depend on a free regulating/interpolation parameter $\eta$. I discuss here how one may fix that parameter using energetics. Other issues discussed are, a linear theory  with an external potential that reproduces  some unusual  exact solutions of the nonlinear equation, and possible symmetry enhancements in the nonlinear theory.

\end{abstract}

\vspace{0.5in}

\section{Introduction}

One purpose of Science is to construct economical theories of observed phenomena. In reality, the data that is available is limited in its accuracy and range because of limitations of technology, time avaliable to collect it, and various uncontrollable fluctuations. In statistical mechanics and quantum theory the fluctuations are described using probabilities, although one usually distinguishes the two fields by saying that the fluctuations in quantum theory are of an intrinsic nature while in statistical physics they are a convenient way of describing a large system.

The concept of entropy, or information, has played a key role in the development of statistical mechanics, and the maximum entropy (uncertainty) method \cite{jaynes} is an elegant way of deducing probability distributions describing a system. It basically exhorts one to make unbiased choices within any given constraints, such as fixed energy. In adopting this procedure one needs a way of quantifying information, or equivalently its inverse, uncertainty. Shannon construted his uncertainty measure by requiring it to satisfy certain plausible axioms \cite{info}; changing those axioms leads to different uncertainty measures and, through the maximum uncertainty method, to different results for the same system.

It turns out that the non-relativistic Schrodinger equation may also be understood within the maximum uncertainty philosophy \cite{reg}: If the initial position of a classical particle is not known precisely, then one adopts a probabilistic approach that leads to classical ensemble dynamics. Simultaneously maximising our uncertainty, by minimising an appropriate ``Fisher" information measure, leads to the Schrodinger equation.

On may motivate the use of the Fisher measure on the basis of plausible axioms in the context of classical ensemble dynamics \cite{par1}, just as the Shannon measure was derived from a slightly different set of axioms appropriate for statsitical mechanics. Yet, the two information measures are not unrelated, being linked by an intermediary called the Kullback-Liebler (KL) distance measure. The Fisher measure arises from the KL measure as a length scale $L$ vanishes while the Shannon measure formally is the infinite $L$ limit of the KL measure \cite{reg,par2}.    

Thus it appears natural to investigate the generalised, nonlinear, Schodinger equation that arises from classical ensemble dynamics when the KL measure is used in the maximum uncertainty procedure. In higher than one space dimension the equation that was obtained is not rotationally invariant, leading to the suggestion that spacetime symmetries might be linked to quantum linearity, with $L$ possibly linked to gravitational effects \cite{par2}.

Consider for simplicity a single particle in one space dimension. The nonlinear equation is 
\begin{equation}
i \hbar {\partial \psi \over \partial t} = - {{\hbar}^2 \over 2m} {\partial^2 \psi \over \partial x^2} + V(x) \psi + 
F(p) \psi \, , \label{nsch1}
\end{equation}
with
\begin{eqnarray} 
F(p) &=& {\cal{E}} \left[ \ln {p(x) \over p(x+L)}  \ + 1 \ - {p(x-L) \over p(x)} \right] + {{\hbar}^2 \over 2m}  {1 \over \sqrt{p}} {\partial^2 \sqrt{p} \over \partial x^2} \label{fip}
\end{eqnarray}
and $p(x) = \psi^{\star}(x) \psi(x)$. The nonlinearity length scale $L$ and the energy parameter ${\cal{E}}$ are constrained through the relation 
\begin{equation}
{\cal{E}} L^2 = {{\hbar}^2  \over 4 m } \, , \label{uncert}
\end{equation}
to ensure that the leading term in the small $L$ expansion of (\ref{nsch1}) yields the usual linear Schrodinger equation.

The nonlinear Schrodinger equation (\ref{nsch1}) resembles a differential-difference equation as the evolution of the wavefunction depends, any fixed time, not just on knowledge at the point $x$ but also at neighbouring points a finite distance away, $x \pm L$. The nonlinearity is also non-polynomial.  Nevertheless, the equation shares a number of important properties with the linear Schrodinger equation, such as the conservation of probability and existence of the usual plane wave solutions.

The equation above should be regularised as there are potential singularities where $p(x)$ vanishes. Let 
\begin{eqnarray}
p_{\pm}(x) & \equiv & p(x \pm \eta L) \,  ,
\end{eqnarray}
where the dimensionless parameter $\eta$ takes values $0 < \eta \le 1$, then the regularised  expression  
\begin{equation}
{ {\cal{E}}  \over \eta^4}  \left[ \ln {p \over (1-\eta) p + \eta p_{+} } + 1 - {(1-\eta) p \over (1-\eta) p + \eta p_{+}} - {\eta p_{-} \over (1-\eta) p_{-} + \eta p} \right] \,  \label{Q2}
\end{equation}
is  to be used in (\ref{fip}) instead of the first term there. Formally $\eta$ also plays the role of an interpolating parameter as one has the fully nonlinear theory at $\eta =1$ and the usual linear quantum mechanics at $\eta =0$.

\section{Perturbative Results and the value of $\eta$}

Schrodinger's equation has been well tested and so any fundamental nonlinearity must be tiny. Thus the 
nonlinearity may be treated as a perturbative correction to the usual linear Schrodinger equation for the purpose of  computing the leading deformation of the energy spectrum for various external potentials. In Ref.\cite{tab} it was found that for smooth external potentials $V(x)$, the energy shift for unperturbed states with nodes is given by a relatively simple expression,
\begin{equation}
\delta E \approx {\frac{\hbar ^{2}|L| \pi}{6m}}\ \sqrt{\eta(1 - \eta)}\ \
(1-4 \eta)
\sum_{p=1}^{N}C_{np}^{2}\, + O(L/a)^2 \label{EPX}
\end{equation}
where $a$ is the characteristic length scale of the linear theory. The coefficients $C_{np}$ depend on the slope of the unperturbed wavefunctions near the nodes, and hence on the external potential. However the dependence of the energy shifts on $\eta$ is universal, resulting in positive  energy shifts for small $\eta$ and negative shifts for larger values. 

So far $\eta$ has been a free parameter, $0 < \eta <1$, describing a family of nonlinear equations. We may also think of $\eta$ as a dynamical variable in ``theory space" and fix it through some argument. One possibility is to see if energy shifts are minimised for some given $\eta$: that is, we assume that the nonlinear theory flows to those values of $\eta$ that minimise the energy of the system. Elementary algebra reveals that the perturbative shifts above indeed reach a unique global minimum at 
\begin{equation}
\eta_m = {7 + \sqrt{33} \over 16} \approx 0.80 \, .
\end{equation}
At this value of $\eta$, the leading energy shifts are {\it negative} so that the nonlinearity reduces the energy of the original linear system.

The result (\ref{EPX}) is for unperturbed states with nodes. For states without nodes we have \cite{tab}
\begin{eqnarray}
\delta E(L) &\propto& L^2 \eta ^{2}\int_{\infty
}^{\infty }{\frac{dx}{p^{3}}}\;  \mbox{[} 6(2-3\eta )^{2}(p^{\prime })^{4}\ -\
12(3-8\eta \ +\ 6\eta ^{2})p(p^{\prime })^{2}p^{\prime \prime }  \nonumber \\
&&\hspace{2cm}\;\;\;+\ 4p^{2}p^{\prime }p^{\prime \prime \prime }\ +\
p^{2}(3(p^{\prime \prime })^{2}-2pp^{\prime \prime \prime \prime }) \mbox{]} \, ,
\label{formal}
\end{eqnarray}
which depends on the wavefunction and hence the external potential. Note that the leading energy shift for such nodeless states is suppressed by an additional small factor $L/a$ compared to the states with nodes. For the simple harmonic oscillator ground state we have explicitly
\begin{equation}
\delta \tilde{E} = {\frac{ \eta^2 ( 1 - \eta ) ( 1 - 3 \eta) }{4 }}
\left(\frac{ L }{a }\right)^2 + O(L^4) \, .  \label{sho-o}
\end{equation}
This has a unique global minimum at $\eta = (3+\sqrt{3})/6 \approx 0.79$ giving again a negative contribution to the energy.  Remarkably this minimum is very close to the universal value for the excited states, $\eta_m \approx 0.80$. 

If the ground state wavefunction of the simple harmonic oscillator wavefunction is used as variational approximation to the nodes-less ground state of other external potentials, then it is plausible that the ground state energy shift is given approximately by (\ref{sho-o}) for { \it all} smooth potentials.

This suggests that $\eta_m \simeq 0.8$ is the universal physical value of the nonlinearity parameter, as it minimises the energy for all states in one space dimension.  

For potentials that are separable in three Cartesian dimensions, such as a particle in a box or the simple harmonic oscillator, the above results still hold as the three-dimensional version of (\ref{nsch1}) is separable \cite{par2}. 

In Ref.\cite{tab} it was noted that if one applied the above quantum mechanical result heuristically to field theory, then the usual high energy divergences of quantum field theory might be moderated by the nonlinearity, which might be an effective description of gravitational effects.  

For other studies of nonlinear Schrodinger equations see, for example, Ref.\cite{others} and references therein.

\section{Exact solutions and Stability}

In Ref.\cite{hai}  a class of exact solutions to the equation were constructed for a particle confined to the half-line, $x>0$,
\begin{equation}
\psi_e(x,t) = C \exp{(-\kappa x)} \ \alpha(x) \ \exp{(-iEt/ \hbar)} \, , \label{ants}
\end{equation}
where $C$ is the normalisation, $\kappa >0$  and  $\alpha(x)$ is a periodic function here taken to be  \begin{equation}
\alpha(x) = \sin( {2 \pi x \over \eta L} ) \, . \label{deff}
\end{equation} 

The energy eigenvalue, $E$, is
\begin{equation}
(1- { E \eta^4 \over  {{\cal E}} } ) = \ln [ 1 + \eta (\gamma -1) ] + {1  \over 1 + \eta (\gamma-1) } \, , \label{energy}
\end{equation}
where 
\begin{equation}
\gamma \equiv \exp{(-2\kappa \eta L)} \, , \label{gam}
\end{equation}
and it is bounded both above and from below for $0< \eta <1$,
\begin{equation}
0 > E \ge { {\cal{E}} \over \eta^4} \left( 1- \ln (1- \eta) - {1 \over (1 -\eta) } \right) \, .
\end{equation}

The lower bound is crucial for the solution to be energetically stable, but as we see, as $\eta \to 1$ the lower bound keeps decreasing without limit.

If we use the results from the previous section that  $\eta_m \sim 0.8$ is the physical value, then we attain  a stable, non-perturbative (in $\eta$), solution to the nonlinear Scrodinder equation. 
This solution appears to be quite unusual: it exhibits a  periodicity (exponentially damped) in the absence of any external periodic potential. Furthermore the wavefunctions are highly degenerate, as one may choose any periodic function $\alpha(x \pm \eta L) = \alpha (x) \, $ and still have the same energy eigenvalue!

\section{A Linear Theory}
Given the unusual nature of the exact solution in the last section, it is of some interest to see what kind of external potential can reproduce such behaviour within the {\it linear} Schrodinger equation. The potential would be given by 
\begin{equation}
V(x) = \left( {E + {\hbar^2 \over 2m} (\partial^2) \over \psi_e} \right) \psi_e \, ,
\end{equation}
with the $\psi_e$ of Eq.(\ref{ants}).
This gives 
\begin{equation}
V(x) = A + B\cot{\beta x} \, ,
\end{equation}
where $A,B,\beta$ are parameters related to those in (\ref{ants}). The singularities of this potential match exactly the nodes of the wavefunction (\ref{ants}) so that the linear Schrodinger equation is well-defined for this case. It might be interesting to study this peridodic, but singular, potential more generally.

\section{Open Issues}  

The leading perturbative correction for excited states discussed in Sect.(2) vanishes at three points $\eta =0,1/4,1$. The point $\eta =0$ is the formal linear limit so that is not surprising. The other two points are intriguing as the theory is nonlinear there. However the $\eta =1$ value corresponds to the unregularised, singular, theory and so probably only the $\eta =1/4$ value is both sensible and interesting. Since the perturbative energy correction vanishes for excited states at those points, one wonders if the theory develops some protective symmetry at those values.

Likewise, the high degeneracy of the exact solution in Sect.(3) suggests that the nonlinear theory might have some interesting hidden symmetries.

In many models of unified theories in high energy physics, extra dimensions are introduced and the full lorentz symmetry is spontaneously broken. Might it be that the lorentz-violating nonlinearity of Ref.\cite{par2} is related to extra dimensions? Several other questions and speculations related to the nonlinear equation are discussed in the cited references.

\section*{Acknowledement}
I thank Robert Conte,  Prosenjit Singha Deo, Sayan Kar, Gelo Tabia and \\
Hai-Siong Tan for stimulating discussions on the above material. I also thank the organisers of the Nonlinearity IV workshop for their hospitality and the opportunity to present our results in a wonderfully relaxing  environment.

\end{document}